\magnification=1200
\baselineskip=20pt
\def\cl{\centerline}

\bigskip
\bigskip
\bigskip

\cl{\bf On the Kraichnan Model of Passive Scalar Advection } 
\cl{\bf Near the Batchelor limit }

\vskip 25pt

\cl{ Alain Pumir }
\vskip -9pt
\cl{Institut Non Lin\'eaire de Nice, 1361 route des Lucioles }
\vskip -9pt
\cl{F-06560, Valbonne, France}

\cl{Boris I. Shraiman}
\vskip -9pt
\cl{Bell Laboratories, Lucent Technologies}
\vskip -9pt
\cl{700 Mountain Avenue, Murray Hill, NJ 07974}

\cl{\&}

\cl{Eric D. Siggia}
\vskip -9pt
\cl{Cornell University, Laboratory of Atomic \& Solid State Physics}
\vskip -9pt
\cl{Ithaca, NY  14853-2501}
\vskip 25pt

\bigskip
\vskip 20pt

\cl{\bf Abstract}
\vskip 20pt
 The third order
correlation function of the scalar field advected by a Gaussian random
velocity, with a spatial scaling exponent $2 - \epsilon$, and in the
presence of a mean gradient, is calculated
perturbatively in $\epsilon << 1$. This expansion corresponds to the
regime close to Batchelor's advection by linear diffeomorphisms. The
scaling exponent is found to be equal to $1$ in dimensions 2 and 3, 
up to corrections smaller
than $ { \cal O } ( \epsilon ) $, implying an anomalous scaling 
of the third order correlation function and the persistence of small
scale anisotropy.
\vskip 30pt

\bigskip
\bigskip
\bigskip
\noindent
PACS numbers : 

\noindent
47.27Gs. - Isotropic turbulence; homogeneous turbulence.

\noindent
47.27Te. - Convection and heat transfer.
 
\vskip 50pt

\vfill\eject

\overfullrule=0pt
\baselineskip=20pt
\pageno=2
\def\ub{\underbar}
\def\cl{\centerline}
\def\ni{\noindent}

\magnification=1200

The investigation of the statistics of the passive scalar field
advected by random flow is interesting for the insight it offers
into the origin of intermittency and anomalous scaling of turbulent
fluctuations. The problem studied in this paper is stated simply by:
 
$$ \partial_t \Theta + (\vec u \cdot \nabla ) \Theta =  \kappa
\nabla^2 \Theta \eqno(1)$$
with the scalar field $\Theta$ forced by the externally imposed
gradient, $g$. It is convenient to subtract out the gradient and study
the fluctuating field, $\theta (r) = \Theta (r) - g r $. It turns out
that even a Gaussian random, but scale invariant, velocity field results
in non-trivial anomalous scaling of the passive scalar structure function,
$< ( \Theta (r) - \Theta (0) )^n >$ for $n>2$. This has been argued by
Kraichnan$^1$, via a plausible closure scheme, for a "white noise" model
where :
$$ < v_a (r,t) v_b(r' , t') > = \delta(t-t') C_{ab} (r-r') \eqno(2a)$$
with :
$$ D_{ab}(r) = C_{ab} (0) - C_{ab}(r)
= D_0 \bigl( (d - 1 + \zeta_v) \delta_{ab} - \zeta_v { r^a r^b \over |r|^2 }
\bigr) |r|^{\zeta_v} \eqno(2b)$$
(where $\zeta_v$ is the scaling exponent and $d$ the space dimension)
which he has introduced some 30 years ago.$^2$ The existence of an
anomalous scaling has been demonstrated explicitely by Gawedski and
Kupiainen$^3$, and Chertkov et al.$^4$
for certain limits of this Kraichnan model and by Shraiman and Siggia$^5$ 
for a
generalized phenomenological
model where temporal correlation of the advecting field is set 
by eddy
turnover. These calculations are based on the so called Hopf equations 
- the stationarity
conditions of the equal-time multipoint correlators. For the white 
velocity case these
can be derived exactly$^{6,7}$ extending the original analysis of the 
2-point function
by Kraichnan$^8$. They have the form:

$$ \eqalign{ { \sum_{ i \ne j }^N }  & \Bigl( D_{ab} (r_i - r_j ) +
 \kappa \delta_{ab} \Bigr) \partial_{r_i}^a \partial_{r_j}^b
<\theta(r_1)...\theta(r_N)>  \cr
= & { \sum_{i \ne j}^N } g_a g_b C_{ab} (r_i - r_j)
<\theta...>_{ij}^{N-2} \cr
- & 2 { \sum_{i\ne j}^N } g_a D_{ab} ( r_i - r_j ) \partial_j^b
<\theta...>_i^{N-1}\cr} \eqno(2c)$$
(with implicit summation over repeating indices).
We restrict ourselves to the inertial range of scales, where $r$ is
large enough so that the molecular diffusivity can
be neglected: $r >> \eta \equiv (\kappa / D_0)^{1/\zeta_v}$.

The analysis of Ref.3 is based on the expansion of Eq.(2) in
$\zeta_v << 1$ about the diffusion limit $\zeta_v =0$, while we
consider the complementary limit of $\zeta_v = 2- \epsilon$,
$\epsilon << 1$.   Reality for the white velocity model, $\zeta_v=4/3$
, lies inbetween.
The expansion in small $\epsilon$
is more involved than what was required in Refs 3,4 for two
reasons.  There are an infinite number of degenerate modes
for $\epsilon = 0$ which are all mixed by the perturbation,
which itself is singular$^{5,9}$.   That is the perturbation is formally
small because of $\epsilon$, but in certain restricted regions of
configuration space it is the biggest term in the equation.
It must be treated by the method of matched asymptotic expansions.
The exponent we find for the third order correlator$^{10,11}$
$\lambda_3 \approx 1$ implies
that the anisotropy introduced by the mean gradient, $g$
on the large scales, decays more slowly as one descends in scale
than that predicted by K41 theory$^{12,13}$
(which for
$\zeta_v = 2-\epsilon $ predicts an exponent $1+\epsilon$).
Since the experimental exponent is also approximately one$^{14}$, it
will be of interest to compare also the full coordinate dependence of
the three point correlation
function when the latter becomes available from experiment or
simulations.   One way of expressing our result for this
correlation function, is as an expansion in the degenerate modes
of the $\epsilon=0$ problem.  Our matching determines all the
coefficients explicitly.

Determination of the anomalous exponents reduces to finding
the zero modes of the linear operator entering the Hopf
equation,$^{3,4,5}$ i.e. the left hand side of Eq.(2) which in the present 
model (and $\kappa \rightarrow 0$ limit) is the generalized Richardson 
diffusion operator:
$  L (d, \zeta_v) \equiv
{ \sum_{ i \ne j }^3 }    D_{ab} (r_i - r_j ) 
\partial_{r_i}^a \partial_{r_j}^b $.
The $\zeta_v = 2$ case is the Batchelor limit$^{15}$ which
is constrained by an overall
$SL(2)\times SO(d)$ symmetry
so that the spectrum of $L_0 \equiv L(d,2)$, also refered to here 
as the Batchelor-Kraichnan operator,
can be completely constructed with the help of Lie algebraic methods$^{5,16}$. 
Here it will serve as a starting point for the perturbation theory
for $ \zeta_v = 2 - \epsilon$: 
$$ L(d,  2 - \epsilon) = L_0(d) - \epsilon L_1 (d) \eqno(3)$$
to the leading order in $\epsilon <<1$. 
(Note, that we are ultimately interested in the physical case of
$\zeta_v = 4/3$.)
The perturbation expansion around $\zeta_v = 2$ is a singular problem
which however can be addressed by the method introduced in Ref.9, 
as we explain now.

Let us start with $L_0$.
It is convenient to introduce the variables
$ {\vec \rho }_1 = ( {\vec r }_1 - {\vec r }_2 ) /\sqrt{2}$ and 
${\vec \rho }_2 = ({\vec r }_1 + { \vec r }_2 - 2 {\vec r }_3 )/\sqrt{6}
$
(On occasion we shall refer to $i=1,2$-index labelling the $\rho$ 
vectors as the "pseudo-space" index to distinguish it from the
d-dimensional real space.)
Next we "factorize":
$ \rho_i ^a = { \sum \atop {i'} } R_{i i'} ( \chi ) \xi_{i'} \eta_{i'}^a $, 
where $R$ represents pseudo-space rotations by $\chi$, and 
$\hat \eta_{1,2}$ are 
two orthogonal unit vectors. In $d = 3$, we also define 
$\hat \eta_3 \equiv \hat \eta_1 \wedge \hat \eta_2 \, = \,
\vec \rho_1 \wedge \vec \rho_2  / |  
\vec \rho_1 \wedge \vec \rho_2  | $ each component of which is invariant
under the action of $SL(2)$$^{9}$. Another important invariant
is the area of the $\vec r_1, \vec r_2, \vec r_3$ triangle: 
$\zeta \equiv | {\vec \rho  }_1 \wedge {\vec \rho }_2 | = \xi_1 \xi_2$.

The zero modes of $L_0(d)$ for $d=2,3$ have been constructed
in Ref. 5,9; e.g. in $d = 3$, the complete set of eigenfunctions
has the form:
$$ \psi^{\lambda}_{\nu,q,l,m,m'} ~ = 
e^{i q \chi} \zeta^{\lambda \over 2}
P_{\nu}^{q,m'}(\xi) D^l_{m,m'} ~ (\hat \eta) \eqno(4),$$
where
$\xi \equiv (\xi_1^2+\xi_2^2 )/ 2 \xi_1 \xi_2  $,
$D^l_{m,m'}~ (\hat \eta)$ is the matrix element of the representation of the 
$SO(3)$ group$^{17}$  of order $l$
and $P_\nu^{q,m'}$ is the Jacobi function.$^{18}$ 
(Note that quantum number $m'$ corresponds to rotations
of the $\hat \eta$ triad about $\hat \eta_3 $ in pseudo-space.)
To ensure analyticity in the $
\zeta =  |\rho_1 \wedge \rho_2 |  \rightarrow 0$ limit
(which corresponds to all 3 points of the correlator being on one 
line), $\lambda/2 - max(\nu , -\nu-1) $ must be a positive integer.  This
is because as $\zeta \rightarrow 0$, 
$\,\,\xi \sim \zeta^{-1} \rightarrow \infty$ and
$P_{\nu}^{q,m} (\xi) \sim \xi^{max(\nu , -\nu -1)}$.

The 3-order structure function, or the skewness, which is the physical object
of interest has odd spatial parity and hence is only non-zero in as much
as the mean scalar gradient, $g$, introduces a particular direction.
Hence the relevant eigenfunctions are the p-waves, $l = 1$.
The zero mode of $L_0(d)$ corresponding to
the smallest exponent $\lambda$ is obtained for 
${\lambda \over 2} = \nu $ yielding, in the $l=1$ sector, 
$\lambda = 1$ independent of d. 

We shall need the explicit form of the $L_0$ operator:
$$  {1 \over 2d } L_0 \psi_{l=1}^{\lambda} ( w, \chi , \hat \eta ) \, = \,
\partial_\xi ( (\xi^2 - 1) \partial_\xi \psi ) 
+  { (  {\partial}_{\chi}^2 - I_3^2)  - 2 i \xi I_3 \partial_\chi \over 
4 (\xi^2 - 1) } \psi - \nu (\nu+1)  \psi \eqno(5),$$
where $\nu (\nu +1) \equiv
{d-2 \over 2d }({ \lambda^2 \over 2d} + \lambda ) - {d+1 \over 2d } l(l+d-2)$
and $I_3 \equiv { 1 \over i } (\eta_1 \partial_2 - \eta_2 \partial_1 )$.
In agreement with Eq.(5), the $\partial_\chi$ and $I_3^2$ are diagonalized by
$\exp( iq \chi)$ and $\hat \eta_1 \pm i \hat \eta_2$ the latter corresponding 
to the $l=1, m'= \pm 1$ sector. Requiring the left hand side of
Eq.(5) to vanish would make it
into a Legendre equation$^{18}$ with $\nu$ and hence $\lambda$
entering as an eigenvalue.

Next we define the perturbation operator in Eq.(3): 
$$  L_1 (d) =   {\cal L}+ {d-1 \over 2d} \bigl( l(l+d-2)+ \lambda^2 -d \lambda
-{1 \over d-1} L_0 \bigr)  \eqno(6a),$$
with :
$$ \eqalign { {\cal L } \equiv & \sum_{{\cal S}_3} 
- \ln ( | \rho_1 | )
\times \cr
& \Bigl( (d+1) \rho_1^2 ( \partial_1^a \partial_1^a
- { 1 \over 3 } \partial_2^a \partial_2^a  ) - 2 \rho_1^a \rho_1^b 
( \partial_1^a \partial_1^b - { 1 \over 3 } \partial_2^a \partial_2^b ) 
 \Bigr) } \eqno(6b).$$
In Eq.(6b), the summation extends over all the cyclic permutations of
$ (\vec r_1, \vec r_2, \vec r_3)$, resulting in the 
following symmetry for the $\rho$ :
$ \rho_1 \rightarrow - { \rho_1 \over 2} \pm { \sqrt{3} \over 2} \rho_2$ and 
$\rho_2 \rightarrow
\mp { \sqrt{3} \over 2} \rho_1 - { \rho_2 \over 2 } $. 
The expansion
breaks down for $ - \ln |\rho_1 | << \epsilon^{-1}$.
When $\xi \rightarrow \infty$, {\it i.e.}, when the three points
$\vec r_1$, $\vec r_2$ and $\vec r_3$ are almost aligned, the operator 
$\epsilon {\cal L}$ 
becomes much larger than the Batchelor-Kraichnan operator $L_0$.
This can be seen by expanding the full operator
in the limit $\xi \rightarrow
\infty$. Defining 
$ \psi^{\lambda}_{l=1} ~ \equiv \zeta^{\lambda \over 2} 
\varphi (\xi, \eta , \chi)$, one finds, for d=3 and the $l=1$ sector:
$$ \eqalign { & {\cal L} \varphi = \cr  
& 
- \ln \bigl[1 - (1 - \xi^{-2})\cos(2 \chi) \bigr]
 \times \bigl[ \xi^2 {\cal L }_2 \varphi + \xi {\cal L}_1 \varphi
+ {\cal L}_0 \varphi + {\cal O } (1/\xi) \bigr] \cr
& +  \bigl(\chi \rightarrow \chi + { 2 \pi \over 3 } \bigr)
  +  \bigl(\chi \rightarrow \chi - { 2 \pi \over 3 } \bigr) }
\eqno(7a).$$

 The operators ${\cal L}_i$ are :
$$ { \cal L}_2 \varphi = { 4 \over 3} (\cos^2 \chi - 1) ( 4 \cos^2 \chi - 1)
\Bigl( - ( 2 \xi \partial_\xi - \lambda)^2 + 4 \eta_1 \partial_{\eta_1} \Bigr)
\varphi
\eqno(7b),$$
$$ { \cal L}_1 \varphi = {32 \over 3} \cos \chi \sin \chi (\cos^2 \chi - 1)
\Bigl( ( 2 \xi \partial_\xi - 1) 
(\eta_2 \partial_{\eta_1} - \eta_1 \partial_{\eta_2} ) 
- 2 \eta_2 \partial_{\eta_1} \Bigr) \varphi
\eqno(7c),$$
and :
$$ \eqalign 
{  { \cal L}_0 \varphi & = -{ 2 \over 3 }  
( \cos^2 \chi - 1) (4 \cos^2 \chi -1)  
\bigl( \partial_{\chi^2} + 5 \bigr) \varphi \cr 
+ & { 64 \over 3 } \sin^3 \chi \cos \chi
\partial_\chi \varphi + { 4 \over 3} 
( 1 + 4 \cos^2 \chi - 8 \cos^4 \chi ) \eta_1 \partial_{\eta_1} \varphi }
\eqno(7d).$$
In Eq.(7d), $\lambda$ has been set equal to $1$ - its unperturbed value -
since the corrections would be higher order in $\epsilon$.
The singular nature of the perturbation follows from the
fact that the ${\cal L}_2$ term enters with prefactor $\xi^2$
so that when $\xi >> (1/\epsilon)^{1/2}$, $\epsilon {\cal L} >> L_0$.
This situation calls for the "boundary layer" type matched asymptotic
analysis which we outline below.

 Let us assume $\lambda =1+ \epsilon \delta$,
define the rescaled 'inner' variable 
$z = \epsilon^{1/2} \xi$, and introduce the function 
$\varphi (z, \chi, \eta ) = z^{\lambda \over 2} 
( \phi_1 (z, \chi ) \hat \eta_1 +
i \phi_2 (z, \chi ) \hat \eta_2 ) $. The prefactor is chosen 
to offset the scaling factor $\zeta^{\lambda/2}$ (see Eq.(4)) which vanishes 
for collinear points.  Physics requires that
$\phi_i$ is bounded when $z \rightarrow \infty$. With this change
of variable and functions, the problem can be written, provided $\chi \buildrel
> \over \sim \xi^{-1}$ as :
$$ \biggl( \Bigl( (z^2 \partial_z^2 + 3 z \partial_z) 
+ { 4 \over 9 } z^2 U(\chi) 
\bigl(  (z \partial_z)^2 - \eta_1 \partial_{\eta_1} 
\bigr) \Bigr) + \epsilon^{1 \over 2} \hat L_1 + \epsilon \hat L_2 + ... \biggr)
( \phi_1 \eta_1 + i \phi_2 \eta_2 ) = 0
\eqno(8a),$$
with 
$$ \eqalign{  U (\chi)  = & 
 \Bigl( (\cos^2 \chi - 1) (4 \cos^2 \chi - 1) 
\ln (1 - \cos(2 \chi)) \cr
+ & ( \chi \rightarrow
\chi + 2 \pi /3 ) + ( \chi \rightarrow \chi - 2 \pi/3) \Bigr) } \eqno(8b).$$ 
The operators $\hat L_1$ and $\hat L_2$ can be deduced from a systematic
expansion of the operator in powers of $\epsilon$ starting from Eq.(7a-d).

 The boundary conditions at infinity imply that when $ z \rightarrow \infty$,
the solution is a function of $\chi$ only. By direct substitution, one finds 
that $\phi_1 = 0$ and $\phi_2 = a(\chi)$, where $a(\chi)$ is 
an unknown function, decomposed for convenience as a Fourier series in
$\chi$ : $a(\chi) = \sum_q \hat a_q e^{i q \chi}$. 
When $z \rightarrow 0$, the problem reduces to the unperturbed 
Batchelor-Kraichnan
operator, up to small corrections. In the matching region, defined by 
$z \rightarrow 0$ but $ \xi \rightarrow \infty$, or equivalently,
$ 1 << \xi << \epsilon^{-1/2}$, the $\xi$ dependence of each Fourier
mode in $\chi$, $q$, must match with 
the asymptotic behavior of the eigenmodes of $L_0$ (Eq.(4,5)) which
is best found via their integral representation given in Ref.5,9,17.
One finds that the functions $\phi_{i,q}$ must behave as:
$ \phi_{1,q} = { \epsilon^{1/2} |q| \over 2 z } + 
{ \epsilon ( 1 - q^2 ) \over 4 z^2 } + ... $ 
and $\phi_{2,q} = (1 - \epsilon { q^2 \over 8 z^2 } + ... ) sign(q) $. 
The crossover equation, Eq.(8), can be solved analytically, to the 
leading order
in $\epsilon$, and the imposition of the matching conditions
determines $a(\chi)$ via an eigenvalue equation for $\delta$:
$$ U (\chi) (\partial_\chi^2 + 1) a(\chi) 
+ 6 \delta a = 0 \eqno(9).$$

 The analysis in $d = 2$ can be carried out in a completely similar way.
As in the
$3$-dimensional case, the behavior
for $\xi \rightarrow \infty $ is of the form 
$\phi = a_{2d}( \chi ) \xi^{\lambda /2} + {\cal O} (\xi^{\lambda/2  -1}) $, 
and the 
function $a_{2d}$ is determined by a matching condition. Surprisingly,
the equation determining $a_{2d}(\chi)$, and the correction to the scaling
exponent, $\delta = (\lambda - 1)/\epsilon$ is identical to Eq.(9).

 Before solving Eq.(9), one needs to determine the appropriate boundary 
conditions. Because
the $3$-point correlation function must be odd under 
$ {\vec \rho }_i \rightarrow - {\vec \rho }_i$, implying that 
$a( \chi + \pi ) = - a( \chi )$.
This, together with the periodicity $ a ( \chi + 2 \pi/3) = a( \chi )$,
resulting from the invariance under cyclic permutation of 
$\vec r_1$, $\vec r_2$ and $\vec r_3$, 
implies that $a(\chi + \pi/3) = - a( \chi )$. The limit 
$ \chi \rightarrow 0$ corresponds to the case where $\vec r_1$ and $\vec r_2$ 
come close together : $ |\vec  r_2 - \vec r_1 | << 
| \vec r_3 - \vec r_1 |, | \vec r_2 - \vec r_1 |$.
In this limit, the correlation function must be invariant when $\vec r_1$
and $\vec r_2$ are permuted, implying that $a ( \chi )$ must be even. Since
$ a $ is even near $ \chi = 0 $ and $a$ is antiperiodic with period $\pi/3$,
$ a ( \pi/6 ) = 0$.

 At small, but finite $\epsilon$, Eq.(9) reduces for $\chi \rightarrow 0$,
to :
$$ - \chi^2 \ln ( \chi ) a '' (\chi) + \delta a(\chi) = 0 \eqno(10a).$$
Introducing the change of variables : $ y \equiv - \ln (\chi)$ and 
$f(y) \equiv a(\chi) $, Eq.(10a) reduces to the following (Kummer)
equation :
$$ y ( f'' + f' ) + \delta f = 0 \eqno(10b).$$
The behavior of the solution when $ \chi \rightarrow 0 $ is:
$$f ( y ) \sim y^{-\delta}  ~~~~~~~ {\rm when } ~~~~~ y \rightarrow \infty 
\eqno(10c).$$
This function diverges (goes to zero) when $\delta < 0$ ($\delta > 0$). 

Since $\chi \rightarrow 0$ (for $\xi^{-1} =0$) corresponds 
to $\rho_1 \rightarrow 0$,
the perturbation expansion leading to Eq.(9) is valid only for 
$ y = - \ln \chi << 1/\epsilon $. Hence, to determine the correct
boundary condition as $ y \rightarrow \infty$ the solution of Eq.(9)
must be matched with the "inner" solution describing the correlator
with two points near coincidence. The latter is governed by the
equation derived
directly from Eq.(3) by expanding in $\rho_1 / \rho_2 <<1$ instead
of $\epsilon$ and which is written conveniently in the polar coordinates
$|\rho_1 / \rho_2|^2 \, =   \xi^{-2} + \chi^2 /4$ and 
$\theta = \arctan ( 2/ \xi \chi )$ (restricting here to $d=2$ for simplicity).
The natural radial variable in this "inner" equation turns out to be
$Y = |\rho_1|^\epsilon$. The region of matching with Eq.(10a) corresponds
to $1-Y <<1$ and $\theta =0$. Quite generally the solution near $Y=1$
behaves as $A + B (1-Y)^\alpha$ with $\alpha >0$ required
to keep the solution from diverging. In the matching region $Y \approx
1+ \epsilon \ln ( \chi/2 )$ so that only the constant term, $A$, must be kept 
when computing to the leading order in $\epsilon$. Comparing with
Eq.(10c) one concludes that matching the "inner" solution is only possible for
$\delta = 0$.

For $\delta =0$ the solution of Eq.(9) is
$ a( \chi ) = \sin (\pi/6 - \chi) / \sin(\pi/6) $ for $0 < \chi < \pi /3$
which is continued over the full range of $\chi$ using reflection symmetry
and periodicity defined above. One observes that
$a( \chi)$ has an apparent $| \chi |$ singularity near
$\chi =0$ (and other points related by symmetry) which is regularized
only for $\chi < e^{-1/ \epsilon}$ via the crossover to the "inner"
solution for nearly coincident points $\rho_1 /\rho_2 <
e^{-1/ \epsilon}$ as discussed above.
Note that although to $O( \epsilon )$
there is no correction to the $\lambda =1$ eigenvalue,
the computed eigenfunction is non-trivial: 
it is a superposition of many $\psi_{1,q}$ modes since 
$a_q \sim 1/q $ for large $q$. 
Also note that the calculations in $2$ and 3-dimensions are 
identical and give the same result: $\delta =0 $.

 Thus, the main result of this paper : the scaling exponent of the
$n = 3$ structure function behaves as $ \lambda = 1$, up to corrections 
smaller than $ \epsilon^1 $.
The exponent of the $3$-point correlation function is therefore smaller 
than the "naive" scaling exponent, equal to $ 1 + 
\epsilon$, therefore demonstrating that the behavior of the skewness near
the Batchelor limit of the Kraichnan, white velocity, model is
anomalous. 
Dispersion in the presence of a mean gradient has been shown 
experimentally$^{19,20}$
and numerically$^{21,22}$ to give rise to strong intermittency effects, 
resulting
in a skewness which remains of order 1, independent of the Reynolds number.
As it is the case in real flows, it is interesting to notice that even
for a white noise velocity field, the anisotropy induced at large scales
decays more slowly than predicted by standard phenomenological 
arguments.$^{12,13,23}$
We conclude by mentioning numerical results demonstrating that a large scale 
anisotropy, such as a large scale shear, imposed on a turbulent velocity 
field, may also result in a large anisotropy at small scale$^{24}$, suggesting
also the existence of an anomalous exponent for the $n=3$ structure function
of the velocity field. 

\vskip 5pt 

\ni
{\ub{\bf Acknowledgements }}

AP thanks DRET for support
under contract \# 95-2591A, and EDS was supported by the NSF under
contract \# DMR9121654

\vskip 5pt
\vfill\eject

{\cl{\ub{\bf References }}}
\vskip 5pt
 
\item{[1]} R. H. Kraichnan, Phys. Rev. Lett. {\bf 72}, 1016 (1994).

\item{[2]} R. H. Kraichnan, Phys. Fluids {\bf 11}, 945 (1968).

\item{[3]} K. Gaw\c{e}dzki and A. Kupiainen, Phys. Rev. Lett. {\bf 75}, 3834
(1995).

\item{[4]} M. Chertkov {\it et al.}, Phys. Rev. {\bf E 52 }, 4924 (1995).

\item{[5]} B.I. Shraiman and E.D. Siggia, C. R. Acad. Sci. Paris, {\bf 321},
S\'er. IIb, 279 (1995).

\item{[6]} B. I. Shraiman and E. D. Siggia, Phys. Rev. {\bf E49}, 2912 
(1994).

\item{[7]} R. H. Kraichnan, V. Yakhot and S. Chen, Phys. Rev. Lett. {\bf 75},
240, (1995).

\item{[8]} R. H. Kraichnan, J. Fluid Mech. {\bf 64}, 737 (1974).

\item{[9]} B. I. Shraiman and E. D. Siggia, Phys. Rev. Lett, {\bf 77}, 
2463, (1996).

\item{[10]} A. Pumir, Europh. Lett. {\bf 34}, 25 (1996).

\item{[11]} D. Gutman and E. Balkovsky, preprint, chao-dyn/9604003, (1996).

\item{[12]} S. Corrsin, NACA R \& M 58B1 (1958).

\item{[13]} J.L. Lumley, Phys. Fluids {\bf 10}, 855 (1967).

\item{[14]} R. A. Antonia and C. W. Van Atta, J. Fluid Mech. {\bf 84}, 
561 (1980).

\item{[15]} G. K. Batchelor, J. Fluid Mech. {\bf 5}, 113 (1959).

\item{[16]} See also E.Balkovsky et al., JETP Lett., {\bf 61}, 1021 (1995) ), 
who investigated the 4 point correlator.

\item{[17]} L. Landau and E. Lifshitz, {\it Quantum Mechanics},
Pergamon Press, New-York (1986).

\item{[18]} G. Vilenkin, {\it Representations of groups and special
functions }, Kluwer Acad. Publ., (1991).

\item{[19]} K.R. Sreenivasan, Proc. R. Soc. London {\bf A 434}, 165 
(1991).

\item{[20]} C. Tong and Z. Warhaft, Phys. Fluids {\bf 6}, 2165 (1994).

\item{[21]} M. Holzer and E.D. Siggia, Phys. Fluids {\bf 6}, 1820 (1994).

\item{[22]} A. Pumir, Phys. Fluids {\bf 6}, 2118 (1994).

\item{[23]} While this paper was in preparation, we learned of 
the numerical study of the spectrum of the 3-point Hopf operator for
the Kraichnan model by O. Gat, 
V. S. L'vov, E. Podivilov and I. Procaccia (submitted to Phys. Rev. Lett.).
However their calculation addressed
the $l = 0$ angular sector, which by symmetry does not contribute to the
3-d order structure function.  
It would be interesting to carry out the numerical
diagonalization in the l=1 sector which could be compared with the
anomalous skewness exponents found in Ref. 9 and in the present paper.

\item{[24]} A. Pumir and B. I. Shraiman, Phys. Rev. Lett. {\bf 75 }, 
3114 (1995).

\end